\documentclass[journal]{IEEEtran}
\usepackage{float}
\usepackage{stfloats}
\usepackage{graphicx}
\graphicspath{{figure/}}
\usepackage{subfigure}
\usepackage{cite}
\usepackage{picinpar}
\usepackage{amsmath,amssymb,amsfonts}
\usepackage{url}
\usepackage{flushend}
\usepackage[utf8]{inputenc}
\usepackage[T1]{fontenc}
\usepackage{colortbl}
\usepackage{soul}
\usepackage{multirow}
\usepackage{pifont}
\usepackage{color}
\usepackage{alltt}
\usepackage[hidelinks]{hyperref}
\usepackage{enumerate}
\usepackage{siunitx}
\usepackage{epstopdf}
\usepackage{pbox}
\usepackage{caption}
\usepackage{array}
\usepackage{booktabs}
\usepackage{nomencl}
\usepackage{stfloats} 
\usepackage{diagbox} 
\usepackage[ruled,linesnumbered]{algorithm2e}

\makenomenclature

\begin{document}

\title{Transferring Reinforcement Learning for DC-DC Buck Converter Control via Duty Ratio Mapping: From Simulation to Implementation}

\author
{
	\vskip 1em{Chenggang Cui, \emph{Member, IEEE}, Tianxiao Yang, \emph{Student Member, IEEE}, Yuxuan Dai, \\Chuanlin Zhang, \emph{Senior Member, IEEE}}
	
	\thanks{This work was supported in part by the Program for Professor of Special Appointment (Eastern Scholar) at Shanghai Institutions of Higher Learning, in part by the National Natural Science Foundation of China under Grant 51607111 and Grant 62173221, and in part by Shanghai Rising-Star program under Grant 20QA1404000. \textit{(Corresponding author: Chuanlin Zhang.)}
			
	C. Cui, T. Yang and C. Zhang are with the Intelligent Autonomous Systems Laboratory, College of Automation Engineering, Shanghai University of Electric Power, Shanghai 200090, China (e-mail: clzhang@shiep.edu.cn).}
}

\maketitle
\pagestyle{empty}
\thispagestyle{empty}

\begin{abstract}
Reinforcement learning (RL) control approach with application into power electronics systems has become an emerging topic whilst the sim-to-real issue remains a challenging problem as very few results can be referred to in the literature. Indeed, due to the inevitable mismatch between simulation models and real-life systems, offline trained RL control strategies may sustain unexpected hurdles in practical implementation during transferring procedure. As the main contribution of this paper, a transferring methodology via a delicately designed duty ratio mapping (DRM) is proposed for a DC-DC buck converter. Then, a detailed sim-to-real process is presented to enable the implementation of a model-free deep reinforcement learning (DRL) controller. The feasibility and effectiveness of the proposed methodology are demonstrated by comparative experimental studies.
\end{abstract}


\printnomenclature
\section{Introduction}
\IEEEPARstart{D}{C} microgrid is becoming more and more attractive due to its conspicuous features compared with AC microgrid: simple structure, easy control, strong robustness and environmental friendly. In the future, power electronics-based DC power system will possibly become a dominating form in national grids. Regarding the control issue for the DC power electronics systems,  it is well acknowledged that several common challenges exist: 1) Many nonlinear phenomena such as bifurcations and chaotic behaviour occur in DC-DC converters mainly due to the switching action among all the different topologies of the circuit \cite{li_TIE_2016}. 2) In order to ensure the stability of the system and detect the real-time status, the protection circuit and the sampling circuit are designed respectively, mostly been neglected in the controller design procedure\cite{bottrell_TPE_2013}. 3) Switching frequencies of power converters have been significantly increased to enhance the power density, which implies that the influences of radiated electromagnetic interference (EMI) is more and more serious \cite{zhang_TPE_2019}. Therefore, the problem of regulating power electronics systems with high performance has been a subject of great interest in recent years and various model-driven control strategies have been proposed. However, it should be noted that those above-mentioned factors would lead to the fact that classical model-driven control methods, such as PID control \cite{Yuan_ISNE_2015}, model predictive control \cite{xu_TPE_2019}, sliding mode control\cite{wang_TPE_2019}, composite control \cite{zhang_TSG_2019}, etc. may behave slow dynamic response speed, output waveform distortion or large fluctuation of the circuit states \cite{li_TPE_2020}. 

Aiming to present a more effective stabilization result for power electronics systems, in recent years, intelligent control methods for DC-DC converters have become a trend and aroused numerous attention from both industrial and scientific communities, see for instances, \cite{li_MPSCE_2020,kapat_PE_2020,chaudhary_RSER_2018}, only mention a few. Sketchily, they can be  classified into three categories: intelligent model-driven methods\cite{adibi_IFAC_2019}, data-driven methods \cite{wang_TPE_2020} and hybrid methods\cite{khooban_TEPCI_2020}. As a typical date-driven control strategy, recent literature has shown that deep reinforcement learning (DRL) has been gradually applied to the advanced control issue for power electronics systems\cite{zhao_TPE_2020}. The main idea of DRL is that the agent searches for an optimal policy to make decisions by interacting with the external environment \cite{kiumarsi_TNNLS_2017}. In the literature, \cite{xia_IECON_2020} designs a distributed multi-agent DRL controller for the islanded DC microgrid and demonstrates the effectiveness via simulation. A  DRL controller based on a fuzzy system is proposed in \cite{khooban_TETCI_2020} to increase the stability of the frequency control subsystem in a microgrid. An adaptive data-driven method based on the ADRC strategy is adopted in \cite{gheisarnejad_TPE_2020} to build a programmable grid with a single voltage bus by DC-DC converters. However, these methods are mainly deployed in simulation setup or partly model-based, whilst very few previous contributions have been dedicated to the practical implementation of pure data-driven DRL methodology into a real-life power electronics system.

The transfer of RL algorithms from simulation to implementation is of practical significance from an industrial application point of view. However, as summarized in \cite{zhu_arXiv_2020}, the transfer may cause several new challenges. Specifically, the gap between the simulation and implementation degrades the performance of the trained policies as the models are utilized in the real-life system. Therefore, extensive efforts are conducted to reduce the sim-to-real gap and accomplish more efficient policy transfer. To some extent, in the real-life DC-DC circuit, the sampling circuit, the protection circuit, unmodeled dynamics and other external disturbances can be considered as the main influence factors, which are possible to render a large steady-state regulation error in practices. 

Regarding the transferring issue of RL from simulation to practices, one of the most widely used methods is transfer learning (TL), which utilizes external experience from tasks to alleviate the burden of learning. The application of TL involves various dimensions, such as Reward Shaping, Inter-Task Mapping, Policy Transfer, etc. \cite{da_AIR_2019}. \cite{book_PE_2021} introduces a transfer method from the offline simulation to the online training and inference on real motor drive systems.  A new algorithm called SPOTA to learn a control policy exclusively from a randomized simulator without using any data from the simulator is proposed in \cite{muratore_CRL_2018}. However, regarding the application of the DRL algorithm into practical DC-DC converters, there are very few existing results that can be found in the literature. Resulting from the real-time requirement as 10kHz or even larger in SiC components and system non-linearity even chaotic behaviour, offline trained RL control strategies for DC-DC converters may sustain unexpected hurdles during the sim-to-real transferring procedure.
\begin{figure}[htb]\centering
	\includegraphics[width=8cm]{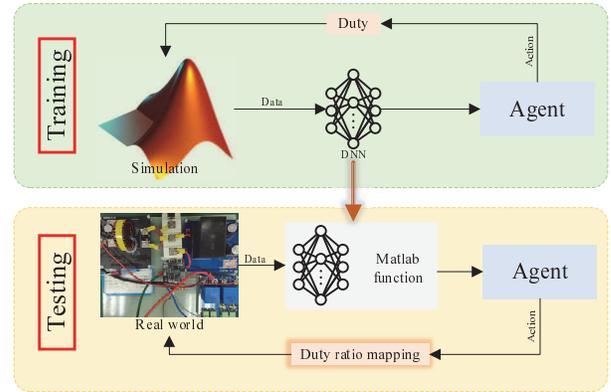}
	\caption{The proposed transferring procedure}\label{introduction}
\end{figure}

As a pioneer work, this paper studies the transferring reinforcement learning for a DC-DC buck converter, aiming to realize the model-free RL controller from simulation to practical implementation. To this aim, as briefly depicted by Fig. \ref{introduction}, a duty ratio mapping (DRM) method is proposed for a DRL control structure in the DC-DC converters. Firstly, a model-free reinforcement learning controller based on the DQN algorithm is adopted to the DC-DC buck converter control with a discrete duty ratio designed as the control actions. Secondly, the DRM is constructed by the voltage conversion ratio under steady-state conditions to transfer the DRL strategy from the simulation to the real-life environment. Thirdly, an approximated linear function is adopted to reconstruct the DRM by measuring the voltage conversion ratio and output current. Finally, the experiment setup is established to evaluate the performance of the proposed transfer approach. The results indicate that the transfer learning using the proposed DRM strategy provides a successful realization to address the challenges of controlling the DC-DC converter with the presence of both internal uncertainties and external variability. The main contributions of this paper can be summarized as the following statements.
\begin{itemize}
\item In order to solve the transfer learning issue for power converters, a new duty-ratio mapping methodology is proposed, which  guarantees the realization of DRL into practices.  
\item Based on the proposed strategy, we are now able to realize the practical implementation of the DRL approach into a DC-DC buck converter while both transient-time and steady-state control performance can be significantly improved in reference to existing related results.
\end{itemize}

The remainder of this paper is organized as follows. Section II describes the basic structure of the DC microgrid with DC-DC converters and the methodology of DRL. In Section III, a new DRM method is proposed to reduce the gap between the simulation and the real DC-DC buck system. Section IV gives the simulations of the buck converter by the proposed DRL controller and the experiments of transfer with the DRM are conducted to demonstrate the validity of the proposed approach. Finally, concluding remarks and future works are presented in Section V.
\section{Problem Formulation and Preliminaries}
\subsection{Problem Formulation}
\begin{figure}[htb]\centering
	\includegraphics[width=0.39\textwidth]{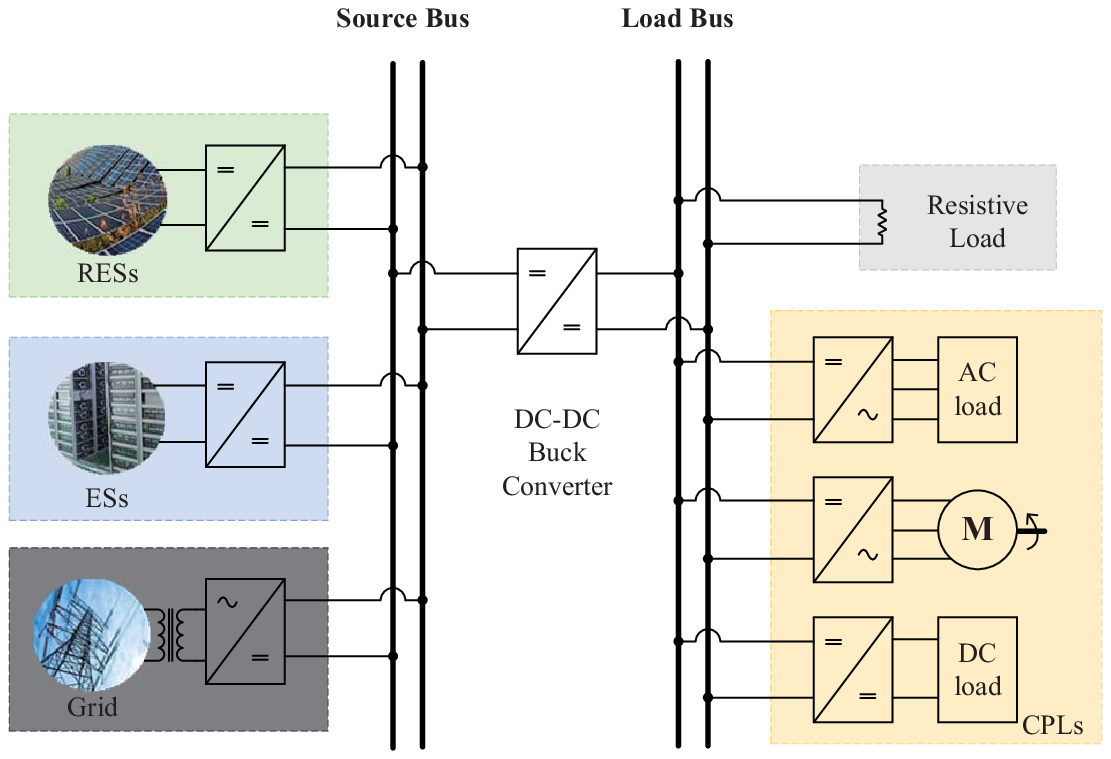}
	\caption{The general layout of a typical DC microgrid}\label{microgrid}
\end{figure}
\begin{figure}[htb]\centering
	\includegraphics[width=0.39\textwidth]{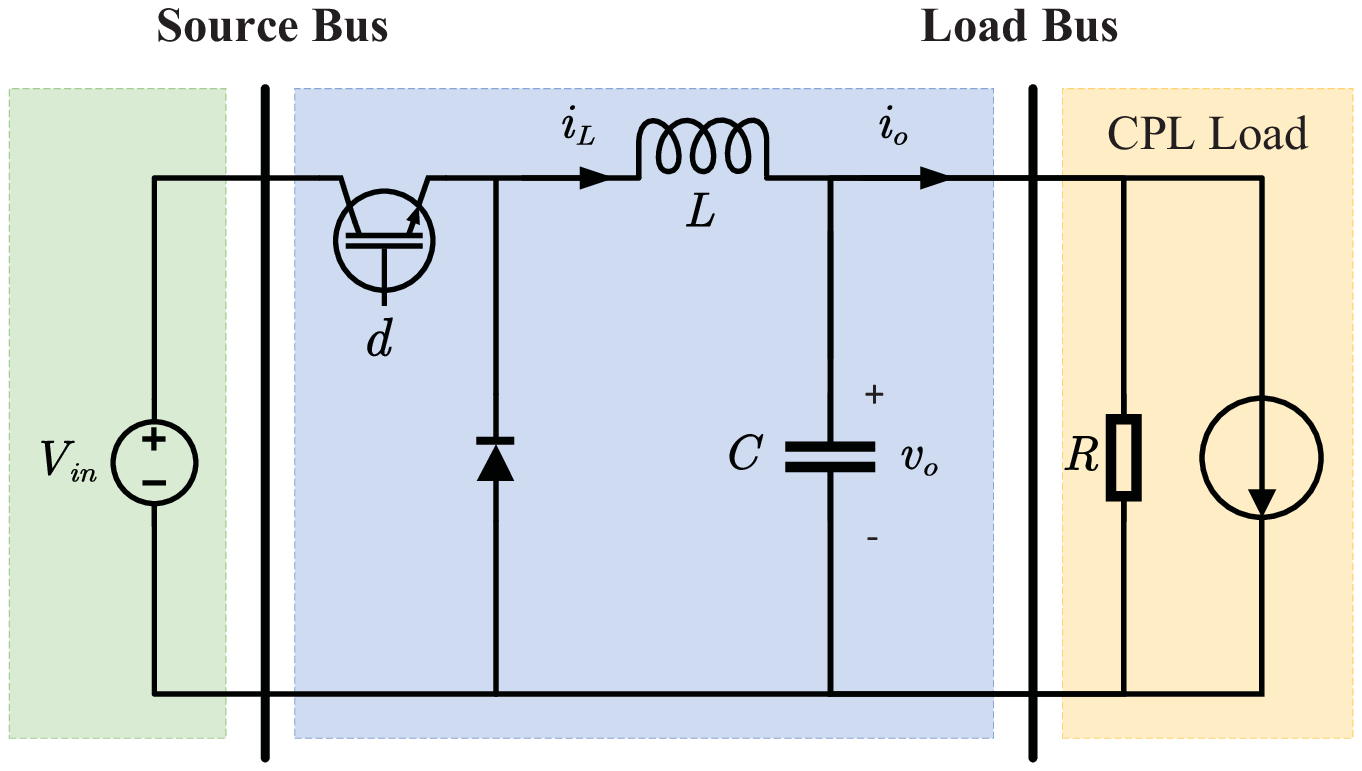}
	\caption{The topology of a DC-DC buck converter}\label{buck}
\end{figure}

Fig. \ref{microgrid} shows a general layout of a typical DC microgrid. Various DC sources including renewable sources (RESs), energy storage systems (ESs) and the AC grid are connected to the source bus. The DC-DC buck converter is adopted to regulate the output voltage at a nominal value for loads. The loads maintain constant power and are supplied by DC load bus via DC-DC converters, DC-AC inverters, which is able to be classified as CPLs. This structure can be found in several similar systems, such as electric vehicles, data centers, electric aircraft and ships, etc. With the determined nominal voltage, the instantaneous current generated by CPLs can be depicted as:
\begin{align}
	i_{\textup{CPL}}=\frac{P_{\textup{CPL}}}{v_{o}}.
\end{align}
\nomenclature{$i_{\textup{CPL}}$}{Current of CPL}
\nomenclature{$P_{\textup{CPL}}$}{Power of CPL}

The minimal conversion unit in the DC microgrid is the DC-DC buck converter shown in Fig. \ref{buck} and its average model is given by \cite{lin_TPE_2019}:
\begin{align}
\begin{cases}
	\dot{i}_{L}=\frac{V_\textup{in}d}{L}-\frac{v_{o}}{L} \\
	\dot{v}_{o}=\frac{i_\textup{L}}{C}-\frac{v_{o}}{R C}-\frac{P_\textup{CPL}}{C v_{o}}.
\end{cases}
\end{align}
\nomenclature{$i_{L}$}{Inductance current}
\nomenclature{$v_{o}$}{Output voltage}
\nomenclature{$V_\textup{in}$}{Input DC source}
\nomenclature{$L$}{Input inductance}
\nomenclature{$C$}{Output capacitance}

The control objective of the DC microgrid is to regulate the voltage of the load bus at a nominal value by the DC-DC buck converter. In this paper, based on a previous DRL control approach in \cite{cui_TCAS_2021}, we are aiming to propose a sim-to-real transferring procedure via a novel DRM strategy. In this regard, both the transient-time and steady-state control performance could be guaranteed. Meanwhile, the accuracy of small-signal and the stabilization of large-signal could be balanced with the proposed control scheme.
\subsection{DRL Methodology Revisit}
 Reinforcement learning is a field of machine learning which researches how to act based on the environment to maximize the expected benefits. The environment is regarded as a Markov decision process (MDP), which gives a tuple as $\{S,A,R,\gamma,P\}$.
\nomenclature{$S$}{States, $s \in S$}
\nomenclature{$A$}{Actions, $a \in A$}
\nomenclature{$R$}{Reward function}
\nomenclature{$\gamma$}{Discount factor, $\gamma \in [0,1]$}
\nomenclature{$P$}{State transition probability}

DQN is a value-based DRL algorithm\cite{bello_arXiv_2016}, which operates according to the maximum Q value in the environment. The key of DQN is a variant of Q-Learning, which inputs raw data and outputs a value function to evaluate the effectiveness of the current action, thereby training the neural network. In the DQN framework, the approximation of the current Q value $y_{j}$ is obtained from a DNN. Each decision will be executed and the Q value will be updated according to the following equation:
\begin{align}
	y_j=\begin{cases}
		r_{j},  \text { if episode terminate at step } j+1; \\
		r_{j}+\gamma \max\limits_{a} \hat{Q}\left(s_{j+1}, a_{j+1} ,  \theta^{-}\right), \text {otherwise.}
	\end{cases}
\end{align}

The gradient descent method is used to reduce the root mean square error of the Q value loss function as much as possible to train the parameters\cite{mnih_nature_2015}. It is depicted as:
\begin{align}
	L(\theta)=E[(r+\gamma \max _{a^{\prime}} Q\left(s^{\prime}, a^{\prime} ; \theta\right)-Q(s, a ; \theta))^{2}].
\end{align}

The DQN algorithm splits the neural network into two parts: 1) One Q network updates the Q value synchronously; 2) The other target Q network calculates the target Q value $y_{j}$, and automatically synchronizes the weight of the Q network to the target Q network after a fixed time step. In the process of interaction between the DQN agent and the environment, the experience data $ \{s_{t}, a_{t}, r_{t}, s_{t+1}\} $ obtained at each time step will be saved to the replay memory $D$, and the neural network will be trained by the batch sampling of the experience data. The action at each step is adopted by:
\begin{align}
	a_t=\begin{cases}
		\arg \max\limits_{a} Q\left(s_{t}, a_{t}\right), & \text { if } p<\varepsilon; \\
		a_{t-\text {ran }} ,                             & \text { otherwise. }
	\end{cases}
\end{align}
\nomenclature{$\varepsilon$}{Exploration rate}
\nomenclature{$a_{t-\text{ran}}$}{Random action}
\nomenclature{$p$}{Random value between 0 and 1 about choosing action}
\subsection{Sim-to-Real Transfer Learning Revisit}
Transferring the DRL strategy from the simulation to practical implementation is necessary for realizing complex real-world engineering applications with RL-based controllers. However, it is not a specific problem of the DRL algorithm but a general problem of all machine learning (ML). Most DRL algorithms provide end-to-end strategies, which are control mechanisms that receive raw sensor data as input and generate direct activation commands as output. The two dimensions of DC-DC converters can be separated similarly. It is of practical significance for simulators to be more accurate to handle the gap between simulation and implementation. Meanwhile, by acknowledging the presence of unmodelled dynamics, this problem can be much more severe, including the general problems of ML to deal with real-world situations that cannot be considered in the simulation.

Therefore, transfer learning is a strategy that takes advantage of the knowledge learned in the source task to facilitate learning the new task goal.  It is mainly utilized to solve the low sampling efficiency and security issues in robotics when the robot or manipulator directly interacts with the environment in the implementation \cite{pham_ICRA_2018}.

\section{Transfer from Simulation to Implementation}
\begin{figure}[htb]
	\centering
	\includegraphics[height=6cm]{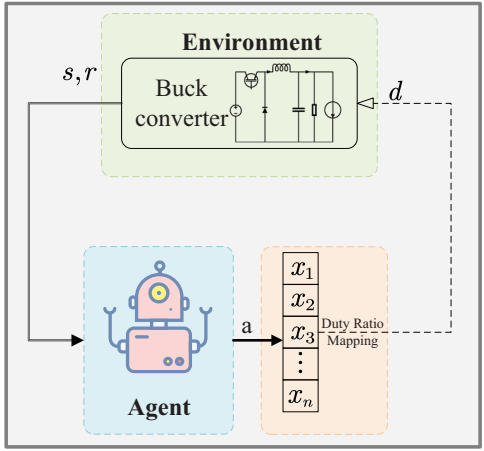}
	\caption{A brief overview of transferring the DRL methodology into real-life systems}\label{Transfer}
\end{figure}

In this section, inspired by the task mapping method, a transfer learning method is adopted to realize the transferring of action-value function for a DRL controller in the real-life environment, shown in Fig. \ref{Transfer}. Detailed design procedures are given as follows.

\subsection{DRL Controller Design}
In a previous work \cite{cui_TCAS_2021}, the authors have proposed a DQN algorithm based on superior learning to adjust the duty ratio for the DC-DC buck converter. The control diagram of the proposed DRL controller is presented in Fig. \ref{structure}. The design of state space, action space, reward/penalty function and exploration strategy is illustrated in the following steps:
\begin{figure}[htb]\centering
	\includegraphics[width=7cm]{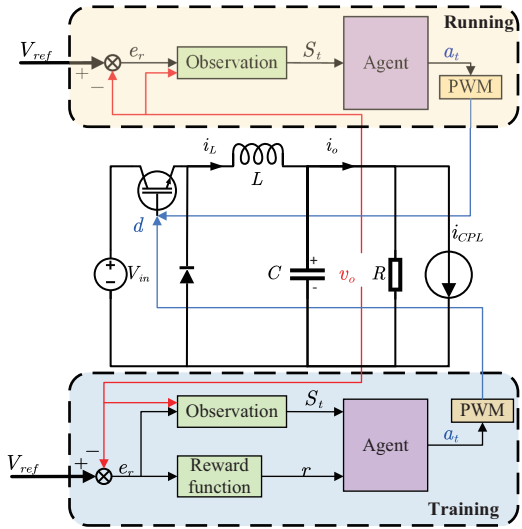}
	\caption{The control diagram of the DRL controller in \cite{cui_TCAS_2021}}\label{structure}
\end{figure}
\nomenclature{$d$}{Duty ratio of the DC-DC buck converter}
\subsubsection{State Space}
The output voltage $v_{o}$ and the tracking error $e(t)=v_{o}(t)-V_{ref}$ are considered as the basic signals to obtain the system state. The state is depicted as:
\begin{align}
	S_{t}=\left\{v_{o}(t), v_{o_{-} \text {del}}(t),  \frac{\mathrm{d} v_{o}(t)}{\mathrm{d} t}, e(t), e_{\text {del}}(t),  \frac{\mathrm{d} e(t)}{\mathrm{d} t}\right\}.
\end{align}
\nomenclature{$V_{ref}$}{Nominal voltage}
\nomenclature{$v_{o_{-}\text{del}}$}{Delay signal of the output voltage}
\nomenclature{$e_\text{del}$}{Delay signal of the tracking error}
\nomenclature{$\frac{d v_{o}(t)}{d t}$}{Time derivative of the output voltage}
\nomenclature{$\frac{d e(t)}{d t}$}{Time derivative of the tracking error}
\nomenclature{$e$}{Tracking error of the output voltage}

\subsubsection{Action Space}
The switch control is chosen as a reference to design a discrete action space. Three variables are defined: the steady-state value $\xi$, the small value $\phi$ to correct the tracking error and the switch signal related to the error feedback $c \in \{-1,0,1\}$. The steady-state value $\xi$ produces a positive or negative deviation. Then, the small value $\phi$ normalize the deviation and the agent can select an effective $c$ to tune the duty ratio quickly. Thereby, a discrete action space is constructed as:
\begin{align}
	A=\{(\xi_1,\phi_1),(\xi_2,\phi_2),...,(\xi_n,\phi_n)\}.
\end{align}

\subsubsection{Reward Function with Sub-goals}
The reward function is designed by the tracking error $e(t)$ between the current state and control objective. Two sub-goals are utilized to guide the learning agent, i.e., $\epsilon_{1}$ and $\epsilon_{2}$. $\beta_{1}$, $\beta_{2}$ and $\beta_{3}$ are selected as the reward/penalty coefficients. The reward function of the proposed controller is shown:
\begin{align}
	r=\begin{cases}
		\beta_{1}-\beta_{3}e(t), & \text { if }0\leq|e(t)|<\epsilon_{1};               \\
		\beta_{2}-\beta_{3}e(t), & \text { if }\epsilon_{1}\leq|e(t)|\leq\epsilon_{2}; \\
		-\beta_{3}e(t),          & \text{else}.
	\end{cases}
\end{align}

\nomenclature{$\epsilon_{1},\epsilon_{2}$}{Subgoals of the expected error}
\nomenclature{$\beta_{1},\beta_{2}$}{Reward coefficients}
\nomenclature{$\beta_{3}$}{Penalty coefficient}

\subsubsection{DNN Design}
The network has seven layers, including an input layer, three fully-connected layers, two hidden layers and an output layer. The hidden layers have both 64 neurons. The activation function of each hidden layer uses the Relu function.

\subsection{Duty Ratio Transfer Functional}
In what follows, a DRM method is introduced in detail as the task mapping construction and the mapping function approximation regarding the DRL control issue for a  DC-DC buck converter.

It is well known that the model in the simulation environment of a DC-DC converter behaves a large deviation from the actual circuit in the real environment. The DRL control strategy stabilises the DC power systems to reach the steady-state value with a high control performance. The task mapping can be constructed  by the parameters of the converters under steady-state conditions, ignoring the dynamic behaviours. Thereafter, the output voltage and the voltage deviation of the DC-DC buck converter conform to the following formulas: 
\begin{align}\label{duty1}
\begin{cases}
	v_{i, \textup{real}}=v_{i, \textup{sim}} ,\\
	v_{o, \textup{real}}=v_{o, \textup{sim}}= v_{\textup{ref}} ,\\
	e_{o,\textup{real}}=e_{o, \textup{sim}}= v_o(t)-v_{\textup{ref}}=0.\\
\end{cases}
\end{align}

\begin{figure}[htb]
	\centering
	\includegraphics[width=0.48\textwidth]{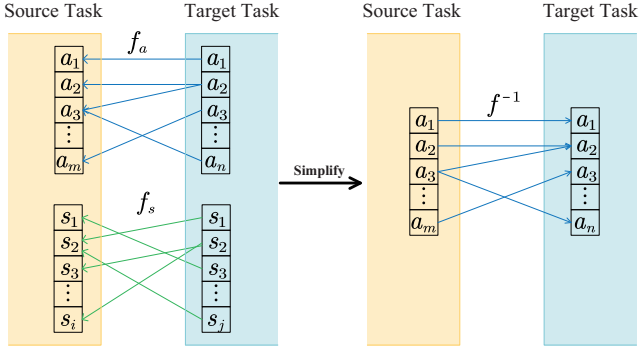}
	\caption{Simplification of the duty ratio mapping (DRM) strategy}\label{S_T}
\end{figure}

For the sake of simplicity, as shown in Fig. \ref{S_T}, we only need to consider the state-action transform relationship between states $v_{o}$ and actions $d$. 

On the one hand, in the simulation environment, the steady-state voltage conversion ratio of the buck converter is equal to the duty ratio, that is,
\begin{align}\label{duty2}
	v_{o, \textup{sim}} = {d_{\textup{sim}}}{v_{i, \textup{sim}}}.
\end{align}

On the other hand, the action-state transformation relationship in the real-life environment can be expressed as:
\begin{align}\label{duty3}
	{v_{o, \textup{real}}} = f(d_{\textup{real}},v_{o,\textup{real}}, i_{o,\textup{real}})v_{i,\textup{real}},
\end{align}
where $f$ is a monotone mapping between the duty ration in the simulation and real-life environment.

According to relations (\ref{duty1})-(\ref{duty3}), the task mapping of the action-state transform can be converted to a duty ratio mapping
\begin{align}\label{duty4}
	d_{\textup{sim}} = f(d_{\textup{real}},v_{o,\textup{real}},i_{o,\textup{real}}),
\end{align}
 i.e.,
\begin{align}\label{duty5}
	d_{\textup{real}} = f^{-1}(d_{\textup{sim}},v_{o,\textup{real}},i_{o,\textup{real}}).
\end{align}
\nomenclature{$f$}{Mapping between the duty ratio in the simulation and real environment}

Thus, a learned DRL control strategy in the simulation environment can be transferred to the real-life environment by creating a DRM given the DC-DC converter simulation environment $D_{\textup{sim}} = (S_{\textup{sim}},A_{\textup{sim}},P_{\textup{sim}},R_{\textup{sim}})$ and the real-life environment $D_{\textup{real}} = (S_{\textup{real}},A_{\textup{real}},P_{\textup{real}},R_{\textup{real}})$.
\nomenclature{$D_{\textup{real}}$}{Information table of experiments}
\nomenclature{$D_{\textup{sim}}$}{Information table of simulations}

\subsection{Mapping Function Approximation}
In real-life experiments, it is necessary to obtain the form of the function approximation for the DRM. Hence, a simple linear function is adopted to approximate the relationship. According to (\ref{duty5}), the DRM is a mapping between the actual duty ratio, simulated duty ratio, and output current under steady-state conditions. Therefore, the sampled data of the DRM can be obtained through the experimental results in the simulation environment and the real-life environment under steady-state operating conditions. Assuming that the actual duty ratio is a linear function of the simulated duty cycle and output current, the approximate function can be regressed by a two-degree linear approximation using a partial least-square method with a set of sampled data.

Referring to Algorithm 1, the detailed process to approximate the DRM is given as follows:
\begin{enumerate}
	\item Given $d_{\textup{real},k}\in A$ in the action set and $P_{o,k}\in[P_{o,\textup{min}},P_{o,\textup{max}}]$ in the constant power load range.
	\item Given the duty ratio action $d_{\textup{real},k}\in A$ and the output power $P_{o,k}$ in the simulation environment, the simulation environment information $D_{\textup{sim},k}(v_{o,\textup{sim},k},d_{\textup{real},k},i_{o,\textup{sim},k})$ with the output voltage and current can be obtained under steady-state operation.   
	\item Given the output power $P_{o,k}$ in the real environment, tune the real duty ratio $d_{\textup{real},k}$ to the simulation duty ratio action $d_{\textup{sim},k}$ under steady-state operating condition. 
	\item The output current $i_{o,\textup{real},k}$ and voltage $v_{o,\textup{real},k}$ of the DC-DC convert of the real environment information $D_{\textup{real},k}(v_{o,\textup{real},k},d_{\textup{real},k},i_{o,\textup{real},k})$ can be measured.	
	\item The simulation duty ratio $d_{\textup{sim},k}$ corresponds to real duty ratio $d_{\textup{real},k}$, which is calculated by (\ref{duty2}), i.e., $d_{\textup{sim},k}=v_{o,\textup{real},k}/v_{i,\textup{sim},k}$.	
	\item Repeat the procedures from 1) to 5). A set of sampled data $S(v_{o,k},d_{\textup{real},k},i_{o,k})$ can be obtained through the experimental results. 	
	\item The approximate function $d_{\textup{real}} = ad_{\textup{sim}}+bi_{o,\textup{real}}+c$ of the DRM is regressed by the partial least square method through multiple groups of transfer sampled data.	
\end{enumerate}
\begin{algorithm}
	\caption{Acquisition of the duty ratio mapping}\label{algorithm}
	\KwData{Output power $P_{o}$, output current  $i_{o,\textup{real}}$, output voltage $v_{o,\textup{real}}$, simulation duty ratio action $d_{\textup{sim}}$}
	\KwResult{Approximate function $d_{\textup{real}} = f(d_{\textup{sim}})$}
	Initialize the experimental equipment\;
	Determine the range of $P_{o}$, $d_{\textup{real}}$\;
	Initialize the information table $D_{\textup{sim}}$ and $D_{\textup{real}}$\;
	$k \gets 0$\;
	\While{$P_{o,k} \in [P_{o,\textup{min}},P_{o,\textup{max}}] $ and $d_{\textup{real},k}\in A$}
	{	Obtain the corresponding information of simulation according to $d_{\textup{real},k}$\;
		Keep the system stable under the given condition $P_{o,k}$\;
		Obtain the output voltage  $v_{o,\textup{real},k}$ and the output current  $i_{o,\textup{real},k}$ under certain working condition\;
		Calculate the parameters of the actual system  $d_{\textup{real},k}$\;
		Update information table $D_{\textup{sim},k}$ and $D_{\textup{real},k}$\;
		Pause equipment and cool down\;
		$k \gets k+1$\;
		
	}
	Obtain the mapping relationship $d_{\textup{real}}$ by least square method.
\end{algorithm}
\begin{figure}[htp]
	\centering
	\includegraphics[width=1\linewidth]{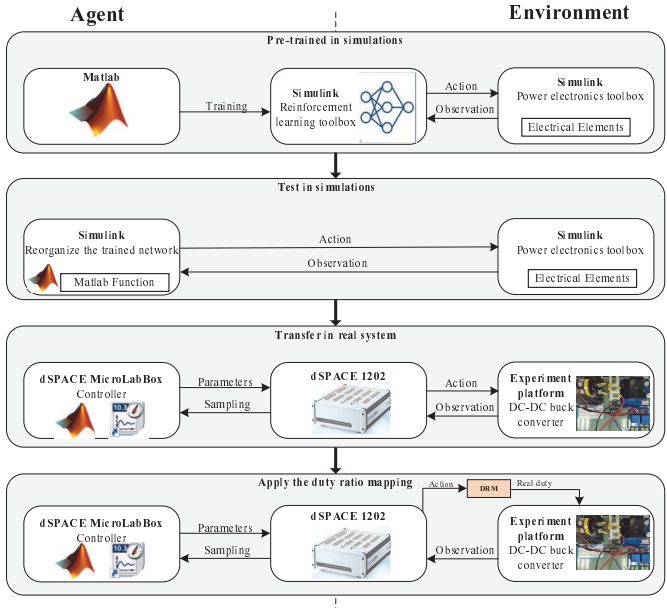}
	\caption{The proposed sim-to-real procedure}
	\label{setup}
\end{figure}
\subsection{Sim-to-Real Procedure}
The development of the proposed controller can be split into four steps, as depicted in Fig. \ref{setup}. Firstly, the power electronics toolbox is utilized to simulate the actual DC-DC converter system. The RL toolbox acts as an intermediary to interact with the environment, and the RL agent’s weights are pre-trained in the simulation. Secondly, the weights are reorganized to synthesize a new network with a Matlab function, which can be regarded as a black box to generate the PWM duty ratio by observation. Thirdly, the compiler exported the controller to C code and imported them to a dSPACE MicroLabBox with a real-time kernel automatically. Then, the controller is running in real-life system without DRM. Finally, the DRM is added in the DRL controller to handle the gap between the simulation and the experimental platform.

\section{Experimental Results and Discussions}
\subsection{Experiment Setup}
\begin{figure}[htb]\centering
	\includegraphics[width=0.9\linewidth]{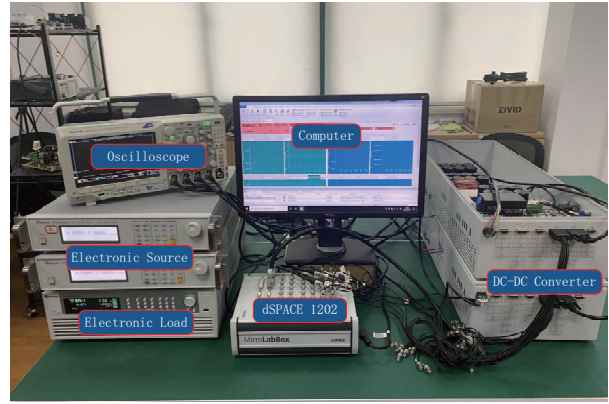}
	\caption{The experiment setup}
	\label{Experimental_setup}
\end{figure}

In this section, the experimental setup depicted in Fig. \ref{Experimental_setup} is built to verify the effectiveness of the proposed transfer strategy, which consists of a custom-designed DC power supply (Chroma 62012P-600-8), a custom-designed DC electronic load (Chroma 63202E-150-200), a DC-DC buck converter and dSPACE 1202. The control algorithm is embedded in dSPACE to generate PWM signals for DC-DC buck converter with a switching frequency of 10kHz. The DC electronic load is configured to operate in constant power mode to simulate the CPL. The detailed parameters of the DC-DC buck converter are shown in Table \ref{parameter_buck}. 
\begin{table}[htb]
	\centering
	\caption{The parameters of the buck converter}	
	\label{parameter_buck}
	\setlength{\tabcolsep}{7mm}{
		\renewcommand{\arraystretch}{1.0}
		\scriptsize
		\begin{tabular}{c c c}
			\hline                                                      
			\bfseries Variables & \bfseries Description      & \bfseries Value \\
			\hline
			$ V_{in} $          & Input voltage    & 200V            \\ 
			$V_{ref}$           & Bus voltage     & 100V            \\ 
			$ \emph{L} $       & Inductance  & 2mH             \\ 
			$ \emph{C}$      & Capacitance & 150$\mu$F       \\ 
			$ f$                & Switching frequency       & 10kHz           \\
			\hline
		\end{tabular}
	}
\end{table}
\begin{table}[htbp]
	\centering
	\caption{DQN learning parameters}
	\label{parameter_DQN}
	\resizebox{\columnwidth}{!}{
		\begin{tabular}{c c c}
			\hline
			\bfseries Variables              & \bfseries Description                         & \bfseries Value      \\ 
			\hline			
			$ \alpha$                        & Learning rate                                 & 0.001                \\ 
			$\gamma$                         & Discount factor                               & 0.9                  \\ 
			$  {B}  $                        & Replay memory capacity                        & 1e-6                 \\ 
			$b $                             & Minibatch size                                & 256                  \\ 
			$\varepsilon$                    & Exploration rate                              & 0.1                  \\ 
			$d$                              & Duty ratio                                    & 0.5\\
			$\beta_{1}, \beta_{2}$ 			 & Reward coefficients                  		 & 
			10, 1   \\ 
			$\beta_{3}$ 					 & Penalty coefficient                  		 & -10\\
			$\epsilon_{1},\epsilon_{2} $     & Sub-goals of the reward function              & 0.1, 1               \\  
			$M$                              & Number of neurons					         & 64\\ 
			$N$                              & Number of neurons 						     & 64\\
			\hline
		\end{tabular}
	}
\end{table}

\begin{figure*}[htbp]
	\centering 
	\includegraphics[width=0.8\textwidth]{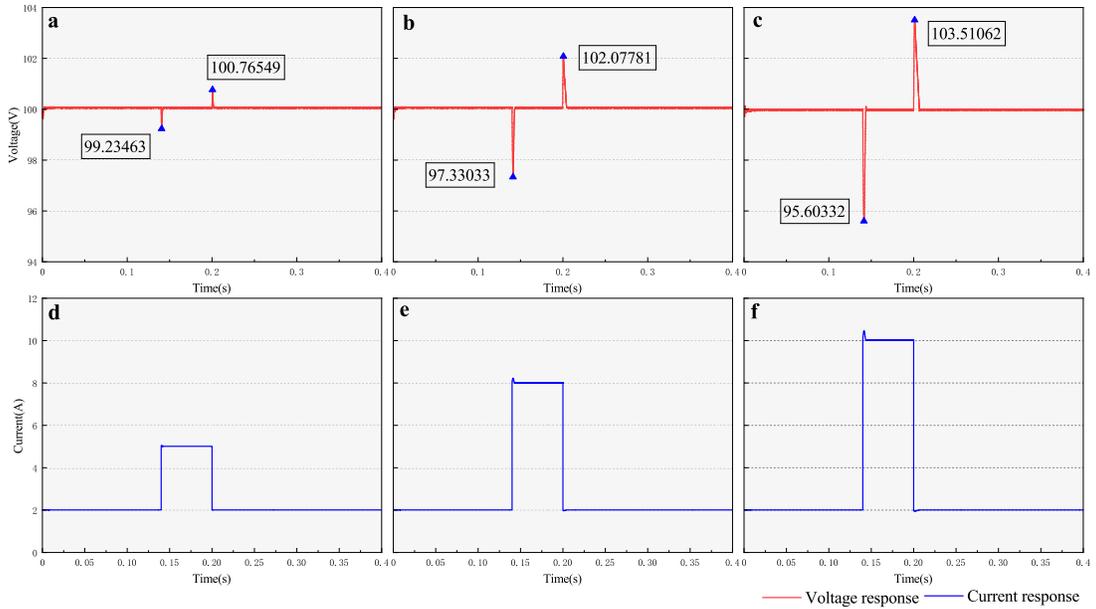}
	\caption{\textbf{Voltage and current response curves of offline training with CPL variations}: \textbf{a and d} from 200W to 500W, and from 500W to 200W; \textbf{b and e} from 200W to 800W, and from 800W to 200W; \textbf{c and f} from 200W to 1000W, and from 1000W to 200W. }
	\label{sim}
\end{figure*}

\begin{figure*}[htbp]
	\centering
	\includegraphics[width=0.98\textwidth]{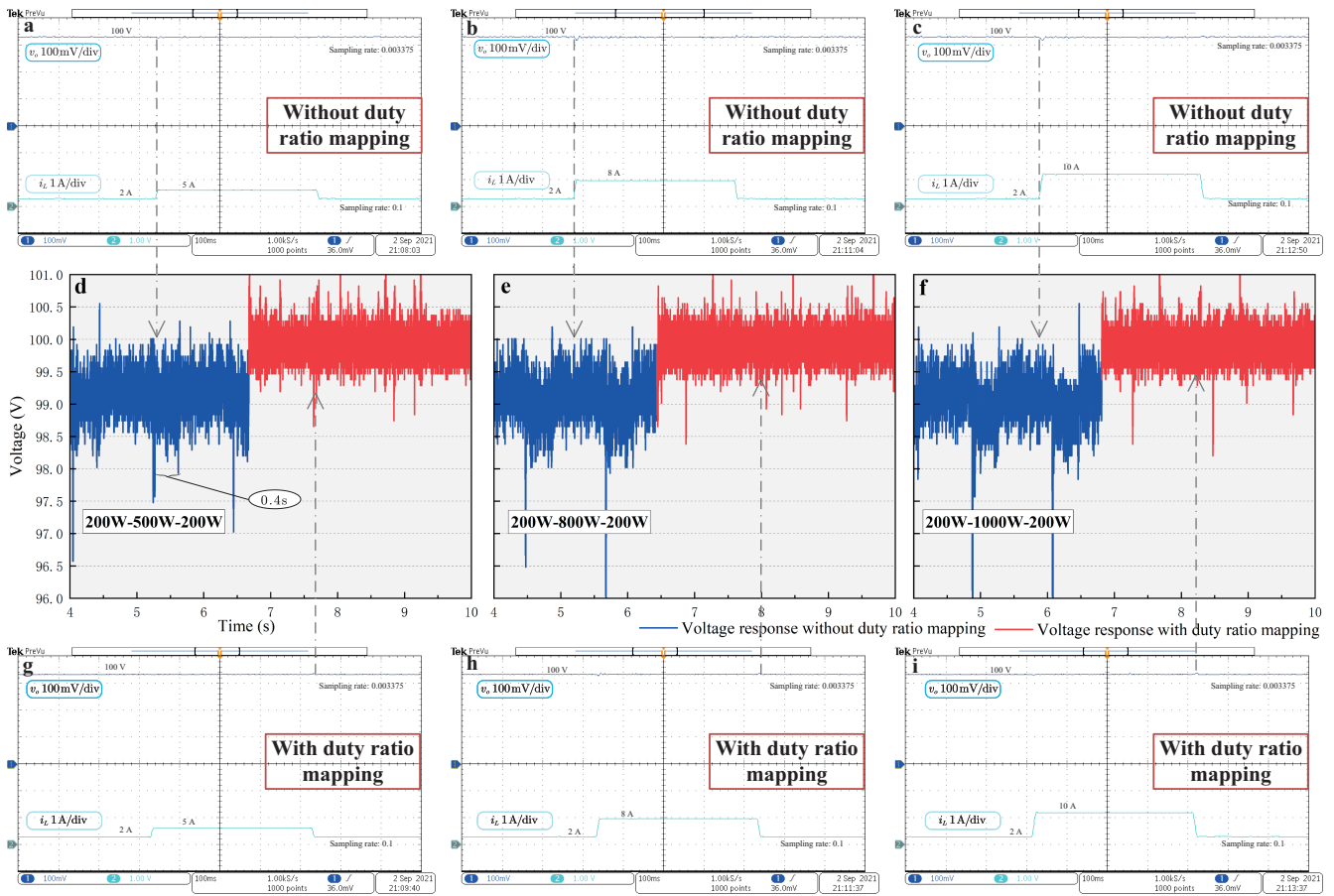}
	\caption{\textbf{Experimental comparison results without and with DRM}: \textbf{a, d and g} from 200W to 500W, and from 500W to 200W; \textbf{b, e and h} from 200W to 800W, and from 800W to 200W; \textbf{c, f and i} from 200W to 1000W, and from 1000W to 200W.}
	\label{Controldesk}
\end{figure*}
\subsection{Offline Training}
The detailed simulation parameters of the DC-DC buck converter is consistent with the real-life system, as shown in Table \ref{parameter_buck}. The parameters of the hyper-parameters for the design of the DQN controller are depicted in TABLE \ref{parameter_DQN}\cite{cui_TCAS_2021}.

In this section, the initial state of the constant power load is set as 200W. Later on, it switches to a new working condition at 0.14s and drops back to 200W at 0.2s. The CPL is switched to the three states of 200W, 500W and 800W, respectively.

As is depicted in Figs. \ref{sim}, no matter which operating condition is switched to, the settling time does not exceed 5ms, and the steady-state voltage can reach the reference voltage 100V. However, it should be pointed out here that the voltage overshoot increases with the increase of load fluctuation respectively.
\subsection{Transferring Experiment}
In what follows, the electronic load is programmed to CPL mode and switches the working conditions every 0.4s. The experiment is conducted at 500W, 800W and 1000W respectively. 

\textit{Case \uppercase\expandafter{\romannumeral1.} Comparison of Simulation and Experiments:}  As is depicted in the simulation, the steady-state voltage under the proposed DRL controller is less than 0.3V. However, it rises to more than 1V in experiments shown in the left part of Fig. \ref{Controldesk}(d)-\ref{Controldesk}(f). To some extent, it is caused by the difference between the simulation and the real-life system. On the other hand, the transient-time control performance in the experiment remains to be the same as in the simulation study, which reflects in the settling time and overshoot, respectively. When the CPL variance occurs, the voltage can be adjusted to reach a steady state within a short settling time, and the voltage overshoot is less than 5V during all experimental procedures. 

\textit{Case \uppercase\expandafter{\romannumeral2.} Comparison with and without DRM}:  Regardless of different CPL variances, the current fluctuations and overshoot are reduced to a certain extent from the comparisons of Figs. \ref{sim}(a)-\ref{sim}(d), Figs. \ref{sim}(b)-\ref{sim}(e) and Figs. \ref{sim}(c)-\ref{sim}(f). Considering that the voltage changes are difficult to observe and the display of the oscilloscope is not obvious, detailed voltage responses in ControlDesk are utilized to compare with and without the impact of the DRM. As shown in Figs. \ref{Controldesk}(d)-(f), the left parts are the voltage responses without transfer and the right parts are the voltage responses with the DRM. Firstly, the steady-state error of the voltage with the DRM is almost similar to simulation and the voltage fluctuation is reduced significantly. Secondly, the overshoot decreases compared to the situation without the DRM. Meanwhile, the maximum overshoot changes average, which increased or decreased suddenly without the DRM.

\textit{Case \uppercase\expandafter{\romannumeral3}. Comparison of Different CPL Variance Rates}: It can be concluded in Figs. \ref{Controldesk} that as the variance increases from 200W to 1000W, the gap between the steady-state voltage without the DRM and the reference voltage is becoming more and more apparent. Meanwhile, the steady-state voltage fluctuates sharply. However, the voltage responses with DRM are always stable in different working conditions.

\section{CONCLUSION}
In this paper, the transfer of  the DRL controller from offline training to implementation is proposed. A DRM method is proposed for a model-free DRL-based DC-DC buck controller transferring to the real world by the voltage conversion ratio. A simple linear function is adopted to approximate the mapping  by the measurement data stream. The presented DRL-based controller is implemented on a typical laboratory hardware system in real-time with a frequency of 10kHz, which is much larger than that in other fields such as robotics. Owing to the fact that the proposed DRM methodology has a strong potential to improve the optimization effect of the DC-DC buck converter, future works can be extended to motion control and microgrid applications, etc. In addition, other transfer methods considering system dynamics will be further studied in order to support more practical implementations.

\bibliographystyle{ieeetr}
\bibliography{DRL}

\end{document}